\documentclass[sigconf]{acmart}

\usepackage{enumitem}
\usepackage{multirow}
\usepackage{textcomp}
\usepackage[inkscapelatex=false]{svg}
\usepackage[capitalise]{cleveref} 
\newcommand{\Uset}{\mathcal{U}}
\newcommand{\Iset}{\mathcal{I}}

\newcommand{\Model}{\mathcal{M}}

\AtBeginDocument{%
  }

\setcopyright{acmlicensed}
\copyrightyear{2018}
\acmYear{2018}
\acmDOI{XXXXXXX.XXXXXXX}
\acmConference[Conference acronym 'XX]{Make sure to enter the correct
  conference title from your rights confirmation email}{June 03--05,
  2018}{Woodstock, NY}
\acmISBN{978-1-4503-XXXX-X/2018/06}

\begin{document}

\title{HoMer: Addressing Heterogeneities by Modeling Sequential and Set-wise Contexts for CTR Prediction}


\author{Shuwei Chen*, Jiajun Cui*, Zhengqi Xu*, Fan Zhang, Jiangke Fan$^\text{\dag}$, Teng Zhang, Xingxing Wang}
\email{{chenshuwei04, cuijiajun02, xuzhengqi02, zhangfan133, jiangke.fan, zhangteng09, wangxingxing04}@meituan.com}
\affiliation{%
  \institution{Meituan}
  \country{Shanghai, China}
}
\thanks{* Equal contribution.}
\thanks{\dag Corresponding author.}

\renewcommand{\shortauthors}{Chen, et al.}

\begin{abstract}

Click-through rate (CTR) prediction, which models behavior sequence and non-sequential features (e.g., user/item profiles or cross features) to infer user interest, underpins industrial recommender systems.
However, most methods face three forms of \textbf{heterogeneity} that degrade predictive performance:
(i) \textbf{Feature Heterogeneity} persists when limited sequence side features provide less granular interest representation compared to extensive non-sequential features, thereby impairing sequence modeling performance;
(ii) \textbf{Context Heterogeneity} arises because a user's interest in an item will be influenced by other items, yet point-wise prediction neglects cross-item interaction context from the entire item set;
(iii) \textbf{Architecture Heterogeneity} stems from the fragmented integration of specialized network modules, which compounds the model's effectiveness, efficiency and scalability in industrial deployments.
To tackle the above limitations, we propose \textbf{HoMer}, a \textbf{Ho}mogeneous-Oriented Transfor\textbf{Mer} for modeling sequential and set-wise contexts.
First, we align sequence side features with non-sequential features for accurate sequence modeling and fine-grained interest representation.
Second, we shift the prediction paradigm from point-wise to set-wise, facilitating cross-item interaction in a highly parallel manner.
Third, HoMer's unified encoder-decoder architecture achieves dual optimization through structural simplification and shared computation, ensuring computational efficiency while maintaining scalability with model size.
Without arduous modification to the prediction pipeline, HoMer successfully scales up and outperforms our industrial baseline by 0.0099 in the AUC metric, and enhances online business metrics like CTR/RPM by 1.99\%/2.46\%.
Additionally, HoMer saves 27\% of GPU resources via preliminary engineering optimization, further validating its superiority and practicality.

\end{abstract}



\maketitle

\section{Introduction}

\begin{figure*}
\centering
\includegraphics[width=1.0\linewidth]{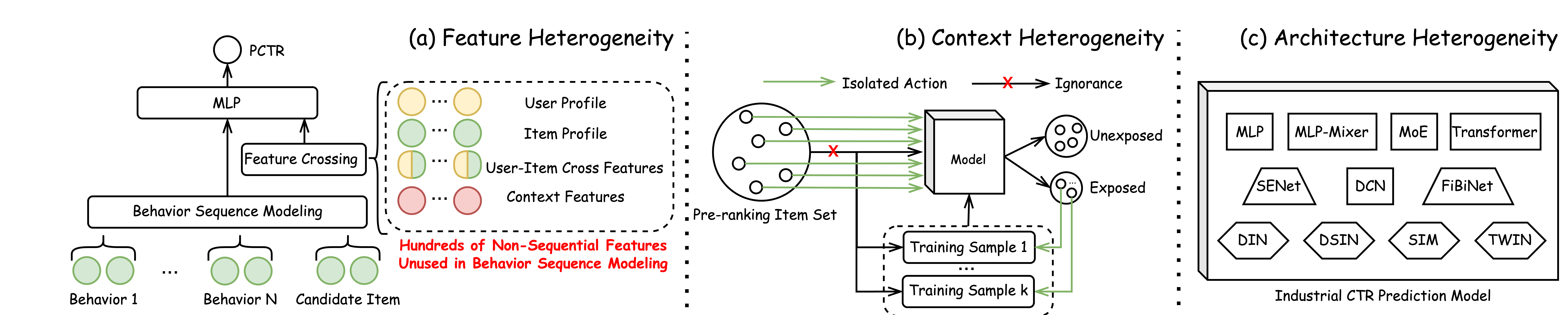}
\caption{
Illustrations of heterogeneities in traditional CTR prediction paradigm.
(a) Feature Heterogeneity: The misalignment between sequence side features and non-sequential features produces coarse-grained user interest representation.
(b) Context Heterogeneity: In point-wise prediction paradigm, the neglect of cross-item interaction context from the entire item set  limits the model's capability to capture authentic user behavior patterns.
(c) Architecture Heterogeneity: The fragmented integration
of specialized network modules constraints the model's effectiveness, efficiency and scalability.
}
\label{fig:motivation}
\end{figure*}

Click-through rate (CTR) prediction refers to the task of estimating the probability of a user clicking on a given item.
It has a direct impact on user engagement and platform revenue, and plays a foundational role in industrial recommender systems.
With the rise of deep learning~\cite{lecun2015deep}, Deep Learning Recommendation Models (DLRMs) have been widely proposed for CTR prediction~\cite{ncf, covington2016deep, dlrm, cheng2016wide, guo2017deepfm}.
Among these developments, behavior sequence modeling~\cite{zhou2018deep, dien, sim, twin, twinv2} and feature crossing~\cite{guo2017deepfm, lian2018xdeepfm, wang2017deep, wang2021dcn, song2019autoint} are two of the most popular topics.
Specifically, the former matches items with a user's behavior sequence to extract dynamic user interest representation, where several side features (e.g., item category and price) are incorporated to enhance sequence representation capability~\cite{fdsa, nova, dif}.
The latter characterizes user interest representation by crossing extensive non-sequential features, including user/item profiles, user-item cross features and context features.
Recent years have witnessed significant advancements in these methods, substantially enhancing CTR prediction performance.

However, we contend that three persistent forms of heterogeneity adversely affect CTR prediction performance:
(i) \textbf{Feature Heterogeneity}: 
As illustrated in~\cref{fig:motivation}(a), both behavior sequence and non-sequential features are utilized to infer CTR.
The extensive non-sequential features are capable for representing fine-grained user interest toward an item, but each behavior is only represented by several side features, producing coarse-grained interest representation.
(ii) \textbf{Context Heterogeneity}: As demonstrated in ~\cref{fig:motivation}(b), the upstream pre-ranking stage generates a set of items (typically numbering in tens to hundreds). The CTR prediction model subsequently predicts CTR for these items in a point-wise paradigm, after which the top-ranked items will be exposed to users on the same display page.
However, a user's interest in any item may be influenced by other co-exposed items within the same page view~\cite{cim, extr}.
For instance, when multiple similar items are exposed concurrently, users tend to engage in comparative evaluation before making click decisions.
The isolated prediction paradigm fundamentally disregards the cross-item interaction context, thereby creating a misalignment with authentic user behavior patterns.
(iii) \textbf{Architecture Heterogeneity}: As shown in ~\cref{fig:motivation}(c), industrial CTR prediction models undergo continuous iteration and integration across multiple technical dimensions, resulting in architectures that inherently contain heterogeneous modules.
For example, DIN~\cite{zhou2018deep}, DSIN~\cite{dsin} and TWIN~\cite{twin} for behavior sequence modeling; SENet~\cite{hu2018squeeze} and FiBiNet~\cite{FiBiNET} for attention-based gating; and DCN~\cite{wang2017deep} and MoE~\cite{shazeer2017outrageously} for feature enhancement.
These modules exhibit overlap or a seesaw effectiveness, with some even becoming ineffective.
Besides, engineering optimization for a single module is unlikely to significantly enhance overall model efficiency.
All these factors complicate model maintenance and iteration, not to mention its potential for scaling up~\cite{hstu, zhou2025onerec, zhou2025onerecv2, rank_mixer, wang2025scaling} in the era of computing.

 We propose \textbf{Ho}mogeneous-Oriented Transfor\textbf{Mer} (\textbf{HoMer}), which models sequential and set-wise contexts in a computationally efficient manner, to tackle  the above heterogeneities.
First, HoMer aligns sequence side features with non-sequential features for fine-grained user interest representation.
It means that each behavior now contains complete user/item profiles, user-item cross features and context features recorded on the corresponding historical request.
For convenience, this paper refers to the constructed sequence as \textbf{panoramic sequence}, as it incorporates all features within the user lifecycle that are perceivable during the CTR prediction stage.
Second, traditional CTR prediction methods construct one sample for each item, and invoke model prediction for each sample in isolation.
HoMer shifts the prediction paradigm from point-wise to set-wise by aggregating the non-sequential features of all the items into one sample, which enables it to perform cross-item interaction and parallel prediction for all items in a single model invocation.
Third, HoMer adopts a unified homogeneous-oriented encoder-decoder architecture~\cite{attention}, where the encoder is responsible for extracting fine-grained user interest representation from the panoramic sequence, and the decoder is responsible for cross-item and user-item interactions.
Under the above settings, HoMer
 ingeniously tackles the three forms of heterogeneity, and the computational overhead introduced by the panoramic sequence is naturally shared by all the items, keeping HoMer's efficiency.

Moreover, HoMer also shows great practicality from multiple dimensions.
Traditional methods generally construct one offline sample for each exposed item, resulting in redundant storage for behavior sequence.
Instead, HoMer constructs one offline sample for each request, which not only reduces storage cost, but also saves data I/O during model training.
To prevent overfitting and reduce training cost, traditional methods usually performs negative sampling on the isolated exposed corpus.
However, this approach inherently hinders the model from understanding rich and authentic user behavior patterns.
By contrast, HoMer is capable of efficiently comsuming the entire training corpus to model more precise patterns, thereby improving prediction performance.
HoMer also simplifies online prediction service.
For example, HoMer's single-pass processing of panoramic sequence inherently eliminates the need for deduplication mechanisms in embedding lookup operations and associated computations.

We conduct extensive offline and online experiments in the search advertising scenario of Meituan\footnote{One of China's largest platforms providing local lifestyle services.}.
The experimental results demonstrate that HoMer outperforms our industrial baseline by 0.0099 in the Area Under Curve (AUC) metric.
It should be noted that a 0.001 AUC improvement is considered significant and worth deploying in industrial recommender systems.
Deployed across Meituan's online traffic from tens of millions of users, HoMer achieves a 1.99\% lift in CTR and a 2.46\% increase in Revenue Per Mille (RPM).
Notably, even with only preliminary kernel fusion optimizations, HoMer has elevated online Model FLOPs Utilization (MFU) from 7.8\% to 12.2\%, while simultaneously reducing online GPU resource consumption by 27\%.
Furthermore, the streamlined architecture and computational efficiency of HoMer enable seamless scalability, yielding significant performance improvements (as discussed in \cref{sec:exp}).

The key contributions of this paper are summarized as follows:
\begin{itemize}[leftmargin=*]
    \item \textbf{Motivation:} We identify three heterogeneity forms (feature, context, architecture) that degrade CTR prediction performance.
    \item \textbf{Methodology:} We propose HoMer, an unified, efficient and practical transformer, to tackle the above heterogeneities.
    It jointly models panoramic sequence for fine-grained user interest representation, and set-wise cross-item interaction context for authentic user behavior patterns learning.
    \item \textbf{Experiment:} We conduct extensive experiments in the search advertising scenario of Meituan.
    Both offline and online experimental results not only consistently validate HoMer's superiority and efficiency, but also clearly demonstrate its scaling potential.
\end{itemize}
\section{Related Work}

\subsection{Deep Learning Recomendation Models}
Deep learning has become the cornerstone of industrial CTR prediction, with models continually advancing in their capacity to capture feature interactions and user behavior pattern. The Deep Learning Recommendation Model (DLRM)~\cite{naumov2019deep} pioneered the integration of embedding layers and multilayer perceptrons (MLPs) to represent complex relationships among features. Building on this, DeepFM~\cite{guo2017deepfm} and Wide \& Deep~\cite{cheng2016wide} combined factorization machines and linear models with deep neural networks to capture both low- and high-order interactions, balancing memorization and generalization. xDeepFM~\cite{lian2018xdeepfm} further improved modeling capabilities via explicit and implicit interaction mechanisms.
Afterwards, user behavior modeling remains central to CTR prediction. DIN~\cite{zhou2018deep} introduced attention-based mechanisms to extract user interests from historical actions, while DIEN~\cite{zhou2019deep} leveraged recurrent networks to capture the evolving nature of user interests. Self-attention techniques, as in AutoInt~\cite{song2019autoint}, have also proven effective for modeling behavior patterns and context. Recent works like Wukong~\cite{zhang2024wukong} highlight the scalability of large unified models, and CIM~\cite{cim} enhances user modeling by incorporating candidate set information.
Despite their successes, these models mostly adhere to a point-wise paradigm and have grown increasingly complex through iterative development, resulting in industrial systems that are challenging to optimize and scale.

\subsection{Transformer in Recommendation}

Transformer architectures have recently gained prominence in recommendation systems due to their strength in modeling complex user-item interactions and long-range dependencies. Their flexibility supports both discriminative and generative recommendation tasks.
Recent research has explored generative paradigms powered by Transformers, aiming to accelerate inference and optimize auto-regressive generation in large-scale scenarios, as seen in EARN~\cite{yang2025earn} and Act-With-Think~\cite{wang2025actwiththink}. Unified frameworks like OneRec~\cite{zhou2025onerec} integrate retrieval and ranking for end-to-end recommendation, while generative pretraining techniques~\cite{wang2025scaling} have improved industrial performance.
Efforts to enhance generative recommenders include advanced tokenization methods, such as ActionPiece~\cite{hou2025actionpiece} and learnable item tokenization~\cite{liu2025generative}, which support context-aware and end-to-end generative modeling. Research on unifying retrieval and ranking~\cite{zhang2025killing} and supporting multi-behavior recommendation~\cite{liu2024multibehavior} further expand the scope of Transformer-based models. EAGER~\cite{wang2024eager} demonstrates holistic list optimization, while collaborative semantics~\cite{zheng2024adapting} and sparse-dense representations~\cite{cobra} showcase the paradigm’s scalability.
Notably, HSTU~\cite{hstu} pushes the boundaries with trillion-parameter sequential transducers, and the generative retrieval paradigm~\cite{tiger} is reshaping the integration of retrieval and ranking. 
Nowadays, transformer-based models support joint inference over multiple candidates are mainstream. Our proposed HoMer does not only continue this trend, but also offering a unified, scalable architecture built from the ground up for end-to-end recommendation.

\section{Preliminary}
A CTR prediction model $\Model$ estimates the probability of a user $u \in \Uset$ clicking on an item $i \in \Iset$ using the following features:
\begin{itemize}[leftmargin=*]
    \item \textbf{User Profile} $f_u$ denotes a user's basic attributes, e.g., age, gender and career.
    \item \textbf{Item Profile} $f_i$ denotes an item's static properties (e.g., category, price and rating) and dynamic metrics (e.g., CTR and popularity).  
    \item \textbf{User-Item Cross Features} $f_{u,i}$ captures crossing signals between user $u$ and item $i$, such as historical click/purchase frequency and affinity scores.
    \item \textbf{Context Features} $f_c$ stands for real-time environmental factors influencing user decisions, such as time of day, location and device type.
    \item \textbf{Behavior Sequence} $f_{seq} = \{f_{seq, 1}, f_{seq, 2}, \dots, f_{seq, n}\}$ records a user's $n$ historical interactions with items (e.g., impressions, clicks, and orders), and will be leveraged to model user interest representation.
    In most methods, several side features (e.g., item category and price) are incorporated to each behavior to enhance sequence modeling, i.e., $f_{seq, i} = \{f_{seq, i}^1, f_{seq, i}^2, \dots, f_{seq, i}^s\}$, where $s$ denotes the number of side features.
    Since the length of behavior sequence is typically long, to maintain data storage and model efficiency, the inequality \( s \ll \lvert f_u \rvert + \lvert f_i \rvert + \lvert f_{ui} \rvert + \lvert f_c \rvert \) generally holds in point-wise prediction paradigm.
    It indicates that sequence side features provide less granular interest representation compared to the non-sequential features.    
\end{itemize}

Following the above notations, the point-wise CTR prediction can be formulated as:
\begin{equation}
    p_{u,i} = \Model(f_u, f_i, f_{u,i}, f_c, f_{seq}),
    \label{eq:point_wise}
\end{equation}
where $p_{u,i}$ is the predicted CTR of user $u$ on item $i$.
This paper proposes a set-wise prediction  method, HoMer, which can be formulated as:
\begin{equation}
p_{u, i_1}, \dots, p_{u, i_k} = \text{HoMer}(f_u, f_c, f_{pan\_seq},f_{i_1},f_{u,i_1},\dots,f_{i_k},f_{u,i_k}),
\label{eq:set_wise}
\end{equation}
where $k$ is the number of items in the set, and $p_{u,i_k}$ represents the predicted CTR of the $k$-th item.
The term $f_{pan\_seq}$ refers to our proposed panoramic sequence, designing to learn fine-grained user interest representation.
The terms $f_{i_k}$ and $f_{u,i_k}$ correspond to the item profile and user-item cross festures of the $k$-th item, which are responsible for modeling cross-item and user-item interactions.
As shown in the equation, HoMer performs set-wise prediction within a single model invocation, ensuring computational efficiency.
\section{Methodology}
Our analysis identifies three inherent heterogeneities in traditional point-wise CTR prediction methods.
 This section presents how HoMer addresses these heterogeneities and improves CTR prediction performance.
First, to tackle feature heterogeneity, HoMer transforms the behavior sequence into panoramic sequence by aligning limited side features with non-sequential features, thereby enabling fine-grained user interest representation.
Second, to address context heterogeneity, HoMer shifts the prediction paradigm from point-wise to set-wise, which facilitates modeling of cross-item interactions and authentic user behavior patterns.
Third, to mitigate architecture heterogeneity, HoMer adopts a unified homogeneous-oriented transformer architecture that incorporates the aforementioned optimizations while maintaining efficiency, practicality, and scalability.
The section concludes with implementation details on model training and deployment.

\subsection{Panoramic Sequence}
\label{sec:panoramic_sequence}
\begin{figure}[t]
\centering
\includegraphics[width=1.0\linewidth]{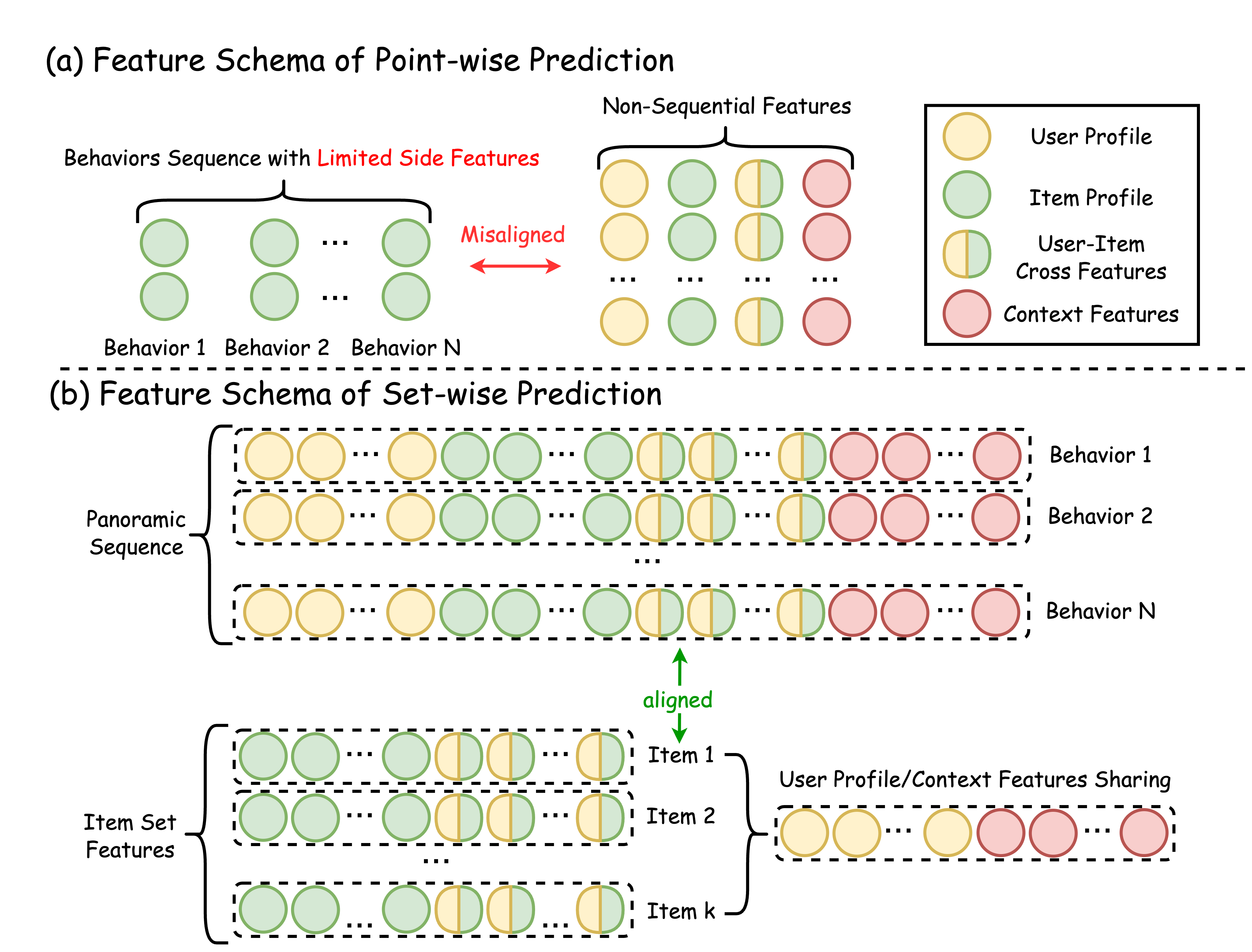}
\caption{
Comparison of feature schemas between point-wise prediction and HoMer's set-wise prediction.
}
\label{fig:feature_schema}
\end{figure}

\cref{fig:feature_schema}(a) illustrates the feature schema of point-wise prediction, which consists of a user's behavior sequence and non-sequential features.
The former contains $N$ behaviors corresponding to $N$ items, each of which is represented by several side features, such as item category and price.
The latter contains extensive features that characterize the user's fine-grained interest representation toward the item.
To capture user's dynamic interest representation toward the item, most CTR prediction methods play matching between the item and the behavior sequence.
However, due to the limited information provided by the several side features, behavior sequence produces coarse-grained user interest representation, resulting in insufficient improvement in CTR prediction performance.

We enrich the behavior sequence by aligning side features with non-sequential features to mitigate feature heterogeneity.
The structure of the resulted panoramic sequence is illustrated in the upper panel of \cref{fig:feature_schema}(b).
For each behavior in panoramic sequence, its side features consist of all the non-sequential features recorded in the corresponding historical request, which can precisely describe the user's decision context at that moment.
In this way, the panoramic sequence contains all the features within the user lifecycle, benefitting behavior sequence modeling and fine-grained user interest representation.
It is important to note that the panoramic sequence can be quickly constructed by chronologically aggregating a user's historical CTR prediction samples.
Moreover, the online log service can be adapted for real-time panoramic sequence construction.

\subsection{Set-wise CTR Prediction}
The point-wise CTR prediction paradigm operates through two distinct phases:
(1) Offline training phase converts impression logs into discrete training samples using strategically sampled negatives; 
and (2) Online serving phase executes independent prediction for individual items.
This presents the following limitations:
\begin{itemize}[leftmargin=*]
    \item \textbf{Storage Inefficiency:} Multiple training samples generated from single page views contain redundant features (identical user profile, context features, and behavior sequence), resulting in inefficient storage utilization, particularly when implementing our proposed panoramic sequence.
    \item \textbf{Sampling Bias:} Negative sampling strategies exclusively utilize exposed items, while systematically excluded unexposed items contain valuable implicit signals for user interest modeling.
    \item \textbf{Contextual Blindness:} The single-item focus of training samples ignores cross-item interaction context within co-exposed page, leading to compromised understanding of authentic user behavior and ultimately constrains model performance.
    \item \textbf{Computational Redundancy:} The serving phase incurs duplicated computations (e.g., table lookup operations and network computations for user profile, context features and behavior sequence) across items from the same request, necessitating deduplication mechanisms to maintain online inference efficiency.
\end{itemize}

\begin{figure*}[t]
\centering
\includegraphics[width=\linewidth]{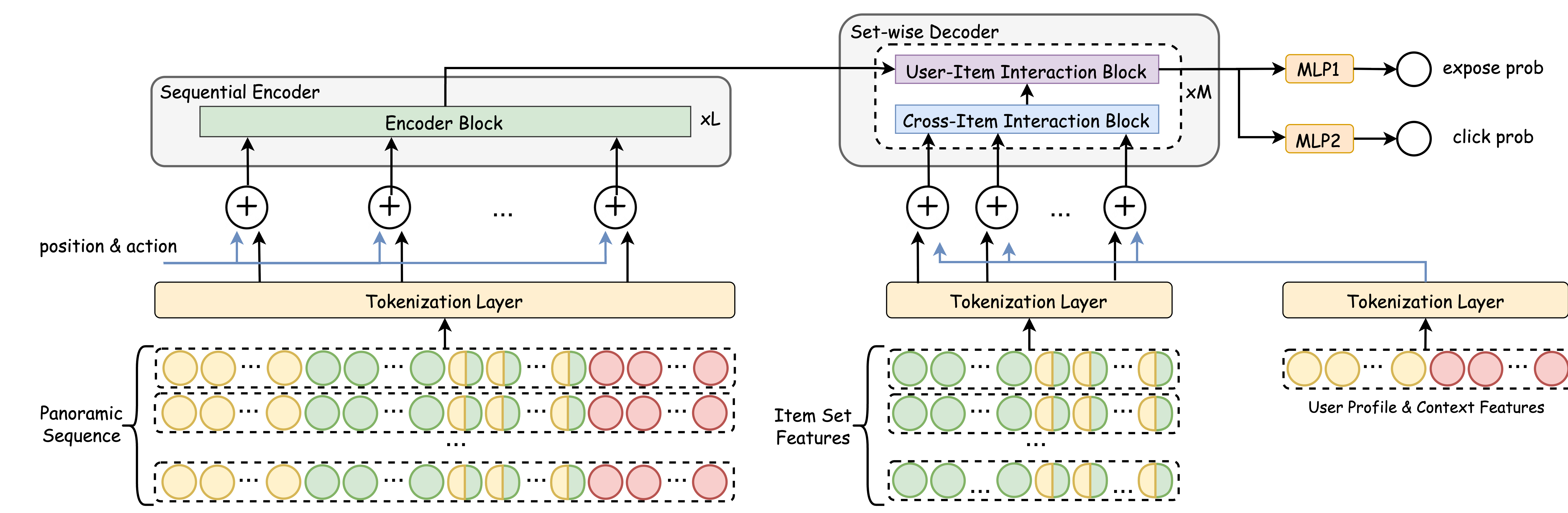}
\caption{
The overall architecture of HoMer. The sequential encoder is responsible for modeling fine-grained user interest representation from panoramic sequence, and the set-wise decoder is tasked with capturing cross-item interaction context from the features of the entire item set.
}
\label{fig:method}
\end{figure*}

In this paper, we introduce a set-wise CTR prediction paradigm designed to address the aforementioned limitations simultaneously.
This paradigm is mathematically formulated in \cref{eq:set_wise}, with the corresponding feature schema depicted in the lower panel of  \cref{fig:feature_schema}(b).
Instead of generating a sample for each item, we construct a single sample for every request.
For items within the same request, their profiles and user-item cross features are aggregated collectively.
Item-independent features, such as user profiles, context features, and panoramic sequence, are shared across all items within the request, thereby reducing storage inefficiency.
In HoMer, items interact with each other to capture authentic user behavior patterns, thus addressing the issues of sampling bias and contextual blindness.
By computing item-independent features only once and sharing them across all items, HoMer facilitates efficient cross-item interaction and parallel prediction for all items through a single model invocation.
Noting that the feature schemas of panoramic sequence and item set are aligned.

\subsection{Homogenous-Oriented Transformer}
Point-wise CTR prediction models typically consist of heterogeneous modules and perform prediction independently for each item.
As discussed in previous sections, this paradigm exhibits inefficiency and suboptimal performance.
In this paper, we introduce HoMer, a unified Homogenous-Oriented Transformer designed to capture both sequential and set-wise contexts while maintaining efficiency, practicality, and scalability. 
\cref{fig:overall} gives an overview of HoMer, which
includes a sequential encoder responsible for modeling fine-grained user interest representation, and a set-wise decoder tasked with capturing cross-item interaction context.
 \subsubsection{\textbf{Sequential Encoder}}
 The sequential encoder is composed of a
series of encoder blocks, and processes panoramic sequence to generate fine-grained user interest representation.
  Each behavior, along with its extensive side features, is passed through a shared tokenization layer and projected into a $d$-dimensional behavior embedding.
After combining these embeddings with the positional and actional embeddings of the behaviors, we obtain the initial user interest representation.
This procedure can be formulated as:
\begin{equation}
    E^0_i=\sigma_1(\phi(f_{pan\_seq,i})) + \sigma_2(\phi(p_i)) + \sigma_3(\phi(a_i))
\end{equation}
where $E^0_i \in \mathbb{R}^d$ is the initial user interest represented by the $i$-th behavior.
Here, $f_{pan\_seq,i}$ refers to the $i$-th behavior and its extensive side features in the panoramic sequence, while $p_i$ and $a_i$ denote the position and user action (e.g., click or not) of the $i$-th behavior.
The term $\phi$ denotes embedding lookup operation, and $\sigma_1$, $\sigma_2$ and $\sigma_3$ are the tokenization layers, which are implemented by a fully connected layer.

Subsequently, the initial representation $E^0 = \{E^0_1, E_2^0, \dots, E^0_N\}$is fed into a stack of encoder block to generate fine-grained user interest representation, which can be formulated as:
\begin{gather}
    Q^l = \tau(\sigma^l_Q(E^{l-1})), K^l = \tau(\sigma^l_K(E^{l-1})), V^l = \tau(\sigma_V^l(E^{l-1})) \\
    E^l = \omega^l(\sigma^l(\tau(Q^l (K^l)^\top) V^l) + E^{l-1})
\end{gather}
where $l \in \{1, 2, \dots, L\}$ is the number of encoder block, $\sigma^l_Q$, $\sigma^l_K$, $\sigma^l_V$, and $\sigma^l$ are fully connected layers within the $l$-th encoder block.
The term $\tau$ refers to the SiLU activation, while $\omega^l$ indicates layer normalization in the $l$-th block.

The fine-grained user interest representation yielded by the final encoder block is then fed into the set-wise decoder to model user-item interactions.
 
 \subsubsection{\textbf{Set-wise Decoder}}
 \label{sec:candidate_decoder}
 The set-wise decoder is composed of a series of decoder blocks.
Each block contains two main components: a cross-item interaction block, which employs the self-attention mechanism to identify dependencies among all items within the current request, and a user-item interaction block, which utilizes the cross-attention mechanism to simultaneously capture the user's dynamic interests across all items.
The input to the set-wise decoder is structured as follows:
Initially, each item's profile and user-item cross features are processed through a shared tokenization layer and projected into a $d$-dimensional item embedding.
Subsequently, the user profile and context features are fed into another tokenization layer and projected into a $d$-dimensional user-context embedding.
The item and user-context embeddings are then combined and fed into the set-wise decoder.
This process can be formulated as:
 \begin{gather}
     H = \bar{\sigma}_1(\phi(f_u) \vert\vert \phi(f_c)), \\
     D^0_i = \bar{\sigma}_2(\phi(f_{i}) \vert\vert \phi(f_{u,i})) + H. 
 \end{gather}
 Here, $H$ is the user-context embedding, and $D^0_i$ denotes the input embedding of the set-wise decoder from the $i$-th item.
 The terms $\bar{\sigma}_1$ and $\bar{\sigma}_2$ are fully connected layers, while $\vert\vert$ indicates concatenation.
Note that the input embeddings of the decoder share similar semantics as the initial user interest representation in the encoder.

Thereafter, $D^0=\{D^0_1, D^0_2, \dots, D^0_K\}$ is fed into a series of decoder blocks for modeling cross-item and user-item interactions, and the computations are conducted as follows:
\begin{gather}
    \bar{Q}^m = \tau(\bar{\sigma}^m_Q(D^{m-1})), \bar{K}^m = \tau(\bar{\sigma}^m_K(D^{m-1})), \bar{V}^m = \tau(\bar{\sigma}^m_V(D^{m-1})), \\
    \bar{D}^m = \bar{\omega}^m(\bar{\sigma}^m(\tau(\bar{Q}^m (\bar{K}^m)^\top) \bar{V}^m) + D^{l-1}), \\
    \hat{Q}^m = \tau(\hat{\sigma}^m_Q(\bar{D}^m)), \hat{K}^m = \tau(\hat{\sigma}^m_K(E^L)), \hat{V}^m = \tau(\hat{\sigma}^m_V(E^L)), \\
    D^m = \hat{\omega}^m(\hat{\sigma}^m(\tau(\hat{Q}^m (\hat{K}^m)^\top) \hat{V}^m) + \bar{D}^m).
\end{gather}
Here, $m \in \{1, 2, \dots, M\}$ is the number of decoder blocks, $\bar{\sigma}^m_Q$, $\bar{\sigma}^m_K$, $\bar{\sigma}^m_V$, $\bar{\sigma}^m$, $\hat{\sigma}^m_Q$, $\hat{\sigma}^m_K$, $\hat{\sigma}^m_V$, and $\hat{\sigma}^m$ are fully connected layers within the $m$-th decoder block.
The terms $\bar{\omega}^m$ and $\hat{\omega}^m$ indicate layer normalization in the $m$-th block.

Finally, the output yielded by the last decoder block is fed into two multilayer perceptrons (MLPs) to estimate the probability of an item being exposed to and clicked by the user:
\begin{gather}
    p^{exp}_i = \text{sigmoid}(\text{MLP}_1(D^M_i)), \\
    p^{clk}_i = \text{sigmoid}(\text{MLP}_2(D^M_i)).
\end{gather}

HoMer integrates both sequential and set-wise contexts, significantly enhancing prediction performance.
It shares user-specific computations across all items, and predicts for them in parallel, showcasing exceptional efficiency.
Besides, HoMer decouples user-specific and item-specific computations, allowing each to be independently scaled up to further improve prediction performance.

\subsection{Training and Deployment}
\subsubsection{\textbf{Training Objective.}}
\label{sec:train_obj}
Traditional CTR prediction models typically utilize cross-entropy loss on the exposed items for training, which can be formulated as:
\begin{equation}
\mathcal{L}_{clk} = -\frac{1}{\vert\mathcal{I}^{exp}\vert} \sum_{i \in \mathcal{I}^{exp}} \left( y_i^{clk} \log p_i^{clk} + (1-y_i^{clk}) \log (1-p_i^{clk}) \right).
\end{equation}
Here, $\mathcal{I}^{exp}$ indicates all items exposed to users, and $y^{clk}_i \in \{0, 1\}$ denotes whether the exposed item is clicked or not.

However, we have observed that relying solely on the loss function defined above may not provide sufficient signals for HoMer to effectively learn complex cross-item interactions.
To address this issue, we apply an auxiliary impression loss, formulated as:
\begin{equation}
\mathcal{L}_{imp} = -\frac{1}{\vert\mathcal{I}\vert} \sum_{i \in \mathcal{I} }\left( y_i^{exp} \log p_i^{exp} + (1-y_i^{exp}) \log (1-p_i^{exp}) \right).
\end{equation}
Here, $\mathcal{I}$ denotes all items generated by the pre-ranking stage, regardless of whether they are exposed to the user.
The term $y_i^{exp} \in \{0, 1\}$ indicates whether the item is exposed to the user.
The final loss combines the above objectives with a balancing hyperparameter $\lambda$, which is set to 1 in this work:
\begin{equation}
    \mathcal{L} = \mathcal{L}_{clk} + \lambda\mathcal{L}_{imp}.
\end{equation}

We use the Adam optimizer~\cite{kingma2014adam} with a learning rate set to $1e^{-4}$ for model training.
Following industry best practices, we employ a one-epoch training strategy, meaning each sample is processed only once to prevent overfitting.

\subsubsection{\textbf{Deployment.}}
In industrial recommender systems, the pre-ranking stage typically generates tens to hundreds of items.
Traditional online CTR prediction services divide these items into multiple shards and perform model predictions for these shards in parallel to improve prediction efficiency.
In our scenario, the 99th percentile of item count does not exceed 300.
By configuring a large shard size, such as 300, HoMer can predict all items in approximately one model invocation.
 When the item count exceeds 300, HoMer must still partition them into several shards.
 Although this may prevent HoMer from fully capturing the entire item set context online, it remains effective as it has already learned cross-item interactions from requests containing fewer than 300 items.

As previously noted, the item count in the majority of requests does not exceed 300.
We utilize Flash Attention~\cite{flashattention} and jagged tensor within our attention-based modules to efficiently manage varying input lengths of the sequential and set-wise contexts.
This approach eliminates redundant computations at padding positions during both the training and serving phases.
Furthermore, due to the structural simplification and homogeneity inherent in HoMer, engineering optimizations applied to the encoder or decoder blocks can effectively enhance the overall efficiency of the model.
By executing kernel fusion on these fundamental blocks, we significantly improve Model FLOPs Utilization (MFU) from 7.8\% to 12.2\%.
Coupled with HoMer's computational efficiency, we achieve a 27\% reduction in online GPU resource consumption.
\section{Experiments}
\label{sec:exp}
In this section, we evaluate our proposed approaches on industrial dataset and aim to answer the following research questions:

\begin{itemize}[leftmargin=*]
    \item \textbf{RQ1:} How does our HoMer model perform in comparison to state-of-the-art DLRMs?
    \item \textbf{RQ2:} What is the improvement in model performance resulting from panoramic sequence modeling and cross-item interaction modeling?
    \item \textbf{RQ3:} How do hyperparameters affect model performance?
    \item \textbf{RQ4:} How does HoMer perform in a real recommender system?
\end{itemize}

\subsection{Experimental Setup}

\subsubsection{\textbf{Dataset}} Our experiments are based on real interaction logs gathered from the search advertising scenario of Meituan.
For offline experiments, we sample logs from April 2025 to July 2025 to build both point-wise and set-wise datasets.
These datasets include 420 million requests from more than 39 million users, featuring nearly 690 thousands unique items and 1.25 billion user behaviors.

\subsubsection{\textbf{Metrics}}
We use \textbf{Log Loss} and Area Under Curve (\textbf{AUC}) as our offline evaluation metrics.
Log Loss is the higher the better and AUC is the lower the better.
Note that a 0.001 improvement in AUC is considered deployment-worthy in our system. For the online A/B testing, we use \textbf{CTR} and \textbf{RPM} (Revenue Per Mille) to measure improvement. Additionally, we examine the \textbf{dense parameter} of the model as well as the \textbf{FLOPs}~(FLoating Point Operations) to align the model's performance under consistent standards. For fairness, We calculate the \textbf{GFLOPs Per Request} for each model. Specifically, each request consists of up to 300 items.

\subsubsection{\textbf{Baselines}}
We implement six commonly used point-wise architectures considering different technical dimensions in industrial recommender systems to compare with HoMer.
To ensure fair comparisons, the panoramic sequence is utilized across all models.

\begin{itemize}[leftmargin=*]
    \item \textbf{Point-wise Baseline.} As shown in Figure \ref{fig:motivation}(c), this baseline involves sequence modeling components (e.g., DIN~\cite{zhou2018deep}), feature interaction components (e.g., DCN~\cite{wang2017deep}), and feature enhancement components (e.g., MoE~\cite{shazeer2017outrageously}).
    \item \textbf{SASRec~\cite{kang2018self}.} This baseline leverages the self-attention mechanism to model sequential dependencies in behavior sequence. 
    \item \textbf{DCNv2~\cite{wang2021dcn}.}  This baseline improves feature interaction learning in recommender systems through enhanced cross-network architecture and low-rank techniques.
    \item \textbf{Wukong~\cite{zhang2024wukong}.} This method introduces a scalable architecture using stacked factorization machines, establishing reliable scaling laws for model performance as complexity increases. 
    \item \textbf{CIM~\cite{cim}.} This method introduces the auxiliary item set features and an impression submodule to improve point-wise  prediction.
    \item \textbf{Point-wise HoMer.} A variant of HoMer that removes the cross-item interaction blocks and $\mathcal{L}_{\text{imp}}$ from the set-wise decoder.
    In this model, although all items are predicted in one model invocation, the items cannot perceive each other.
    To ensure a fair comparison with HoMer, we increase the number of encoder blocks and user-item interaction blocks to maintain a similar FLOPs.
\end{itemize}

\subsection{Overall Performance~(RQ1)}

\begin{figure}[h]
\centering
\includegraphics[width=1.0\linewidth]{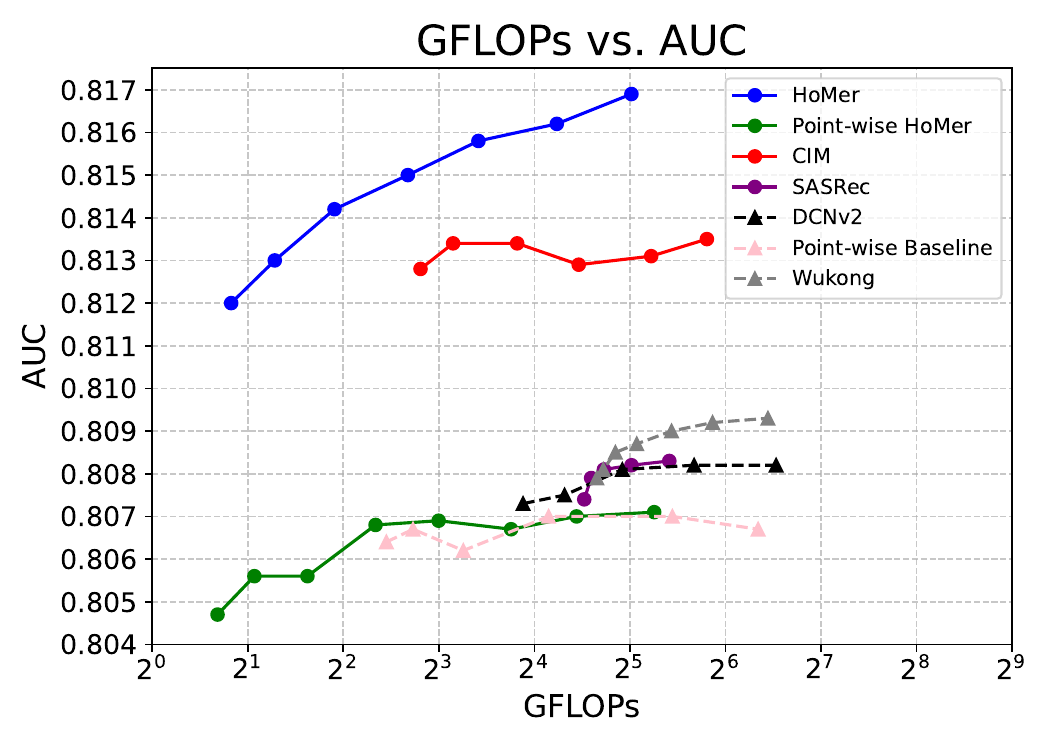}
\caption{
Scalability of models with respect to FLOPs.
}
\label{fig:overall}
\end{figure}

As shown in Table \ref{tab:overall} and Figure \ref{fig:overall}, the proposed HoMer substantially outperforms all competing methods and exhibits a distinct scaling phenomenon.
Compared to other heterogeneous models, HoMer demonstrates a more pronounced performance improvement as FLOPs increases.
We attribute these advantages to the synergistic effects of feauture, context and architecture homogeneities.
Specifically, the performance of point-wise HoMer improves with increasing FLOPs at lower computational budgets, but there exists a threshold beyond which additional scaling yields diminishing returns.
The saturation suggests that it is more beneficial to allocate additional FLOPs to address context heterogeneity.
Moreover, while CIM's base AUC is notably higher due to the modeling of cross-item interaction context, its performance fails to improve with increased model complexity.
We attribute this to the see-saw effect between different network modules introduced by architecture heterogeneity.

\begin{table}[h]
    \centering
        \caption{The main experimental results of HoMer compared with other six baselines.
        The notation $\uparrow$ indicates the higher the better, and the notation $\downarrow$ denotes the lower the better.}
    \label{tab:overall}
\begin{tabular}{c|cc|cc}
\toprule
\textbf{Models}  & \textbf{AUC$\uparrow$} & \textbf{LogLoss$\downarrow$} & \textbf{Flops} & \textbf{Params} \\
\midrule
Point-wise Baseline   & 0.8070       & 0.2464           & 46.8G          & 71.8M           \\
DCNv2            & 0.8082       & 0.2458           & 99.3G          & 159.1M          \\
Wukong           & 0.8093       & 0.2442           & 93.5G          & 120.6M           \\
SASRec           & 0.8081       & 0.2460           & 45.7G          & 58.6M            \\
CIM              & 0.8134       & 0.2438           & 60.1G          & 67.4M           \\
\midrule
Point-wise HoMer & 0.8071       & 0.2464           & 40.9G           & 92.3M           \\
HoMer            & 0.8169       & 0.2425           & 42.6G          & 114.9M        \\ \bottomrule 
\end{tabular}
\end{table}

\subsection{Ablation Study~(RQ2)}

\subsubsection{Effects of Panoramic Sequence Modeling}
In \cref{sec:panoramic_sequence}, we propose a panoramic sequence by aligning sequence side features with non-sequential features.
The side features are categorized into four domains, including user/item profiles, user-item cross features, and context features.
We conduct an ablation study to assess the effects of these side feature domains, with the experimental results encapsulated in  \cref{tab:sideinfo}.
The analysis reveals that each domain contributes positively to HoMer's performance, and their synergistic integration markedly enhances the AUC metric from 0.8128 to 0.8169.
This underscores the notion that these side features harbor complementary information, warranting their comprehensive utilization.

\begin{table}[h]
    \centering
        \caption{Ablation of side features in panoramic sequence.}
    \label{tab:sideinfo}
\resizebox{1.0\linewidth}{!}{
\begin{tabular}{cccc|cc|c}
\toprule
\textbf{User} & \textbf{Item} & \textbf{User-Item} & \textbf{Context} & \textbf{AUC $\uparrow$} & \textbf{LogLoss $\downarrow$} & \textbf{side info \%} \\
\midrule
              &               &    &                 & 0.8128  &  0.2440   & 0.0\%                 \\
               $\checkmark$  &       &    &                    & 0.8149  &   0.2434  & 14.8\%                \\
              & $\checkmark$                      &     &                    & 0.8154    & 0.2432  & 36.3\%                \\
              &               & $\checkmark$     &             & 0.8152   &  0.2431  & 37.9\%                \\
              &               & &  $\checkmark$                & 0.8150   &  0.2433  & 11.0\%                \\
$\checkmark$             & $\checkmark$             & $\checkmark$                  &  $\checkmark$  & 0.8169  &     0.2425 & 100.0\%        \\
\bottomrule
\end{tabular}
}
\end{table}

\subsubsection{Effects of cross-item interaction modeling.} 
\label{sec:abl_candi}
\begin{figure}[!t]
\centering
\includegraphics[width=1.0\linewidth]{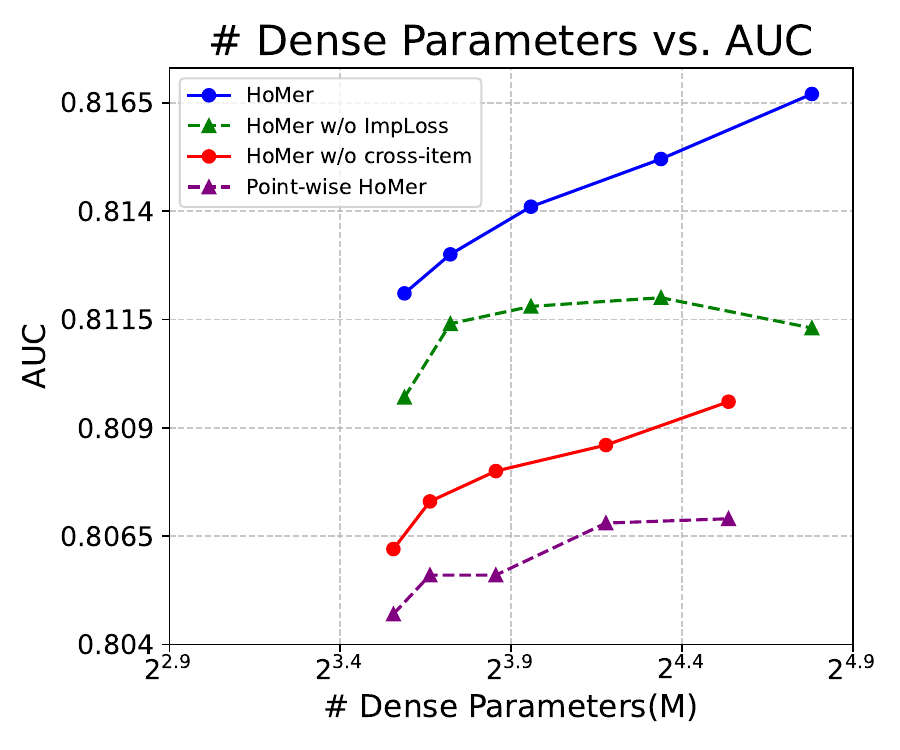}
\caption{
Ablation of cross-item interaction modeling.
}
\label{fig:abl_candi}
\end{figure}

In Section \ref{sec:candidate_decoder} and Section \ref{sec:train_obj}, we delineate the cross-item interaction block and an auxiliary impression loss $\mathcal{L}_{\text{imp}}$, aimed at modeling the contextual relationships among items and capturing authentic user behavior patterns.
To evaluate the performance enhancements attributable to these designs, we perform an ablation study on HoMer and its three variants: point-wise HoMer, HoMer excluding impression loss (HoMer w/o ImpLoss), and HoMer devoid of cross-item interaction blocks (HoMer w/o cross-item). 
Furthermore, we assess the benefits of these designs across varying model sizes by adjusting the number of transformer blocks.
For equitable comparison, we elucidate the correlation between the number of dense parameters in the model and the AUC metric.
As depicted in Figure \ref{fig:abl_candi}, the comparison between the green and red lines with the purple line demonstrates that integrating either the cross-item interaction block or the auxiliary impression loss results in substantial improvements.
The blue and red lines reveal that the inclusion of cross-item interaction context significantly enhances AUC as the model size increases.
A comparison of the blue and green lines indicates that the auxiliary impression loss furnishes effective signals, enabling HoMer to more adeptly learn cross-item interaction context.

\subsection{Hyperparameter Analysis~(RQ3)}
\label{sec:hyper}
We observe that increasing the depth of model, as well as expanding the dimensionality of embeddings and tokens, consistently leads to significant performance improvements.
\subsubsection{Depth of HoMer} 
\begin{figure}[!t]
\centering
\includegraphics[width=1.0\linewidth]{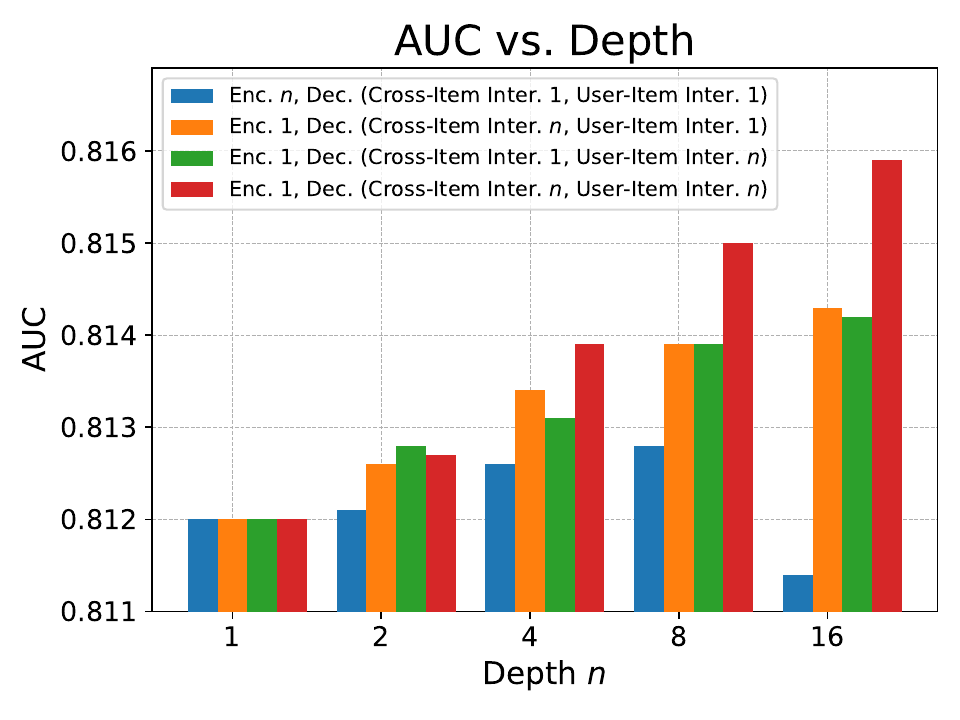}
\caption{
The effects of the depth of HoMer.
}
\label{fig:layer_num}
\end{figure}

\cref{fig:layer_num} illustrates the effects of varying block configurations within the sequential encoder and set-wise decoder.
The blue bars represent configurations with a constant decoder depth while varying the encoder depth, indicating that augmenting solely the number of encoder blocks results in only slight improvements. Conversely, the orange, green, and red bars reveal that both \textit{cross-item interaction blocks} and \textit{user-item interaction blocks} contribute comparably to performance enhancements, and these two blocks complement each other effectively.
These observations imply that HoMer's performance exhibits greater sensitivity to the capacity of the set-wise decoder.

\subsubsection{Dimensionality of embeddings and tokens.}
As presented in Table \ref{tab:dimensions}, increasing the token dimension ($D_{token}$) consistently enhances the model's predictive performance, as reflected by higher AUC values across various layer configurations. For example, when the token dimension is doubled from 256 to 512, the AUC improves from 0.8150 to 0.8162 for $L=8$, $M=8$, and from 0.8161 to 0.8167 for $L=8$, $M=16$. This trend indicates that a larger token dimension enables the model to capture richer and more informative representations, thereby boosting its discriminative capability. On the other hand, increasing the embedding dimension ($D_{embed}$) from 16 to 32 results in only slight improvements in AUC, suggesting that the model's performance is less sensitive to changes in embedding dimensionality compared to token dimensionality. These observations imply that allocating more computational resources to expanding the token dimension is generally more beneficial for improving model accuracy, while increasing the embedding dimension yields diminishing returns. However, it is important to note that higher token and embedding dimensions also lead to increased computational burden, as indicated by the substantial growth in FLOPs and parameter counts. Therefore, a careful trade-off between performance and efficiency should be considered when selecting model dimensions.


   

\begin{table}[!t]
    \centering
     \caption{Prediction performance and computational burden of HoMer with varying model depths, embedding dimensions, and token dimensions.}
    \label{tab:dimensions}
\resizebox{1.0\linewidth}{!}{
\begin{tabular}{c|cc|c|cc}
\toprule
\textbf{Layers \#}          & \textbf{D$_{embed}$} & \textbf{D$_{token}$} & \textbf{AUC $\uparrow$} & \textbf{Flops} & \textbf{Params} \\
\midrule
\multirow{4}{*}{$L=8$, $M=8$}   & 16                & 256               & 0.8150       & 6.9G           & 21.2M           \\
                         & 32                & 256               & 0.8151       & 8.9G           & 35.6M           \\
                         & 16                & 512               & 0.8162       & 20.2G          & 55.8M           \\
                         & 32                & 512               & 0.8164       & 24.3G          & 84.5M           \\
                         \midrule
\multirow{4}{*}{$L=8$, $M=16$} & 16                & 256               & 0.8161       & 11.4G          & 28.8M           \\
                         & 32                & 256               & 0.8163       & 15.4G          & 46.3M           \\
                         & 16                & 512               & 0.8167       & 34.8G          & 79.9M           \\
                         & 32                & 512               & 0.8169       & 42.6G          & 114.9M \\
                         \bottomrule
\end{tabular}
   }
\end{table}

\subsection{Online A/B Test~(RQ4)}

We conduct A/B testing on Meituan’s search advertising scenario with 20\% online traffic between 2025-09-08 and 2025-09-15. 
HoMer achieved gains of CTR+1.99\% and RPM+2.46\% over the highly optimized baseline, validating its superior effectiveness. Now, HoMer has been deployed on the main traffic of Meituan’s search advertising, serving tens of millions of users.

\section{Conclusion}
This study identifies three types of heterogeneities that undermine CTR prediction performance: feature, context, and architecture heterogeneities.
We introduce HoMer, a Homogeneous-Oriented Transformer, to mitigate the aforementioned heterogeneities.
HoMer employs a unified encoder-decoder architecture, with the sequential encoder dedicated to capturing fine-grained user interest representation from the proposed panoramic sequence, and the set-wise decoder focused on modeling cross-item interaction context and learning authentic user behavior patterns.
Comprehensive experiments conducted in Meituan's search advertising scenario demonstrate HoMer's superiority, computational efficiency and scalability.
The online A/B test further underscore HoMer's practical value for large-scale recommender systems.
In future work, we aim to explore engineering strategies to enhance HoMer's scalability, including more efficient training and inference techniques, as well as system-level optimizations, to fully realize its potential in even larger and more complex industrial scenarios.



\bibliographystyle{ACM-Reference-Format}
\bibliography{reference}

\appendix

\end{document}